\documentclass[10pt, aps, prx, twocolumn,  superscriptaddress]{revtex4-2}
\usepackage{setspace}
\usepackage[table]{xcolor}
\usepackage{tabularx}
\usepackage[utf8]{inputenc}
\usepackage{graphicx}
\usepackage{amsmath}
\usepackage{multirow}
\usepackage{booktabs}
\usepackage[]{natbib}
\usepackage{hyperref}
\hypersetup{
  colorlinks=true,
  linkcolor=blue,
  citecolor=blue,
  filecolor=blue,
  urlcolor=blue
}
\usepackage{graphicx}
\usepackage{xcolor}
\setcitestyle{numbers}

\usepackage{eurosym}
\usepackage{tikz}
\usepackage{adjustbox}
\usepackage{amsthm}
\usepackage{upgreek}
\usepackage{booktabs}  
\usepackage{algorithm}
\usepackage{algpseudocode}
\usepackage{enumitem}
\setitemize{noitemsep,topsep=0pt,parsep=0pt,partopsep=0pt,leftmargin=*}
\usepackage{amssymb}

\usepackage{mdframed}
\usepackage{enumitem}

\makeatletter
\renewcommand\@fnsymbol[1]{\relax} 
\makeatother

\begin{document}

\title{Quantum circuits as a game: A reinforcement learning agent for quantum compilation and its application to reconfigurable neutral atom arrays}

\author{Kouhei Nakaji}
\affiliation{NVIDIA Corporation, 2788 San Tomas Expressway, Santa Clara, 95051, CA, USA}
\affiliation{Department of Chemistry, University of Toronto, Lash Miller Chemical Laboratories, 80 St. George
Street, Toronto, ON M5S 3H6, Canada}

\author{Jonathan Wurtz}
\affiliation{QuEra Computing Inc., 1284 Soldiers Field Road, Boston, MA, 02135, USA}

\author{Haozhe Huang$^{*}$}
\affiliation{Department of Computer Science, University of Toronto, Sandford Fleming Building, 10 King’s
College Road, Toronto, ON M5S 3G4, Canada}
\affiliation{Vector Institute for Artificial Intelligence, 661 University Ave. Suite 710, Toronto, ON M5G 1M1,
Canada}

\thanks{$^{*}$ These authors contributed equally.}

\author{Luis Mantilla Calderón$^{*}$}
\affiliation{Department of Computer Science, University of Toronto, Sandford Fleming Building, 10 King’s
College Road, Toronto, ON M5S 3G4, Canada}
\affiliation{Vector Institute for Artificial Intelligence, 661 University Ave. Suite 710, Toronto, ON M5G 1M1,
Canada}

\author{Karthik Panicker}
\affiliation{Department of Chemistry, University of Toronto, Lash Miller Chemical Laboratories, 80 St. George
Street, Toronto, ON M5S 3H6, Canada}

\author{Elica Kyoseva}
\affiliation{NVIDIA Corporation, 2788 San Tomas Expressway, Santa Clara, 95051, CA, USA}

\author{Alán Aspuru-Guzik}

\affiliation{NVIDIA Corporation, 2788 San Tomas Expressway, Santa Clara, 95051, CA, USA}
\affiliation{Department of Chemistry, University of Toronto, Lash Miller Chemical Laboratories, 80 St. George
Street, Toronto, ON M5S 3H6, Canada}
\affiliation{Department of Computer Science, University of Toronto, Sandford Fleming Building, 10 King’s
College Road, Toronto, ON M5S 3G4, Canada}
\affiliation{Vector Institute for Artificial Intelligence, 661 University Ave. Suite 710, Toronto, ON M5G 1M1,
Canada}
\affiliation{Acceleration Consortium, 700 University Ave., Toronto, ON M7A 2S4, Canada}
\date{\today}
\begin{abstract}
We introduce the ``quantum circuit daemon" (QC-Daemon), a reinforcement learning agent for compiling quantum device operations aimed at efficient quantum hardware execution. We apply QC-Daemon to the move synthesis problem called the Atom Game, which involves orchestrating parallel circuits on reconfigurable neutral atom arrays. In our numerical simulation, the QC-Daemon is implemented by two different types of transformers with a physically motivated architecture and trained by a reinforcement learning algorithm. We observe a reduction of the logarithmic infidelity for various benchmark problems up to 100 qubits by intelligently changing the layout of atoms. Additionally, we demonstrate the transferability of our approach: a Transformer-based QC-Daemon trained on a diverse set of circuits successfully generalizes its learned strategy to previously unseen circuits.
\end{abstract}
\maketitle

\section{Introduction}
\label{section:introduction}
\begin{figure*}
    \centering
    \includegraphics[width=0.95\linewidth]{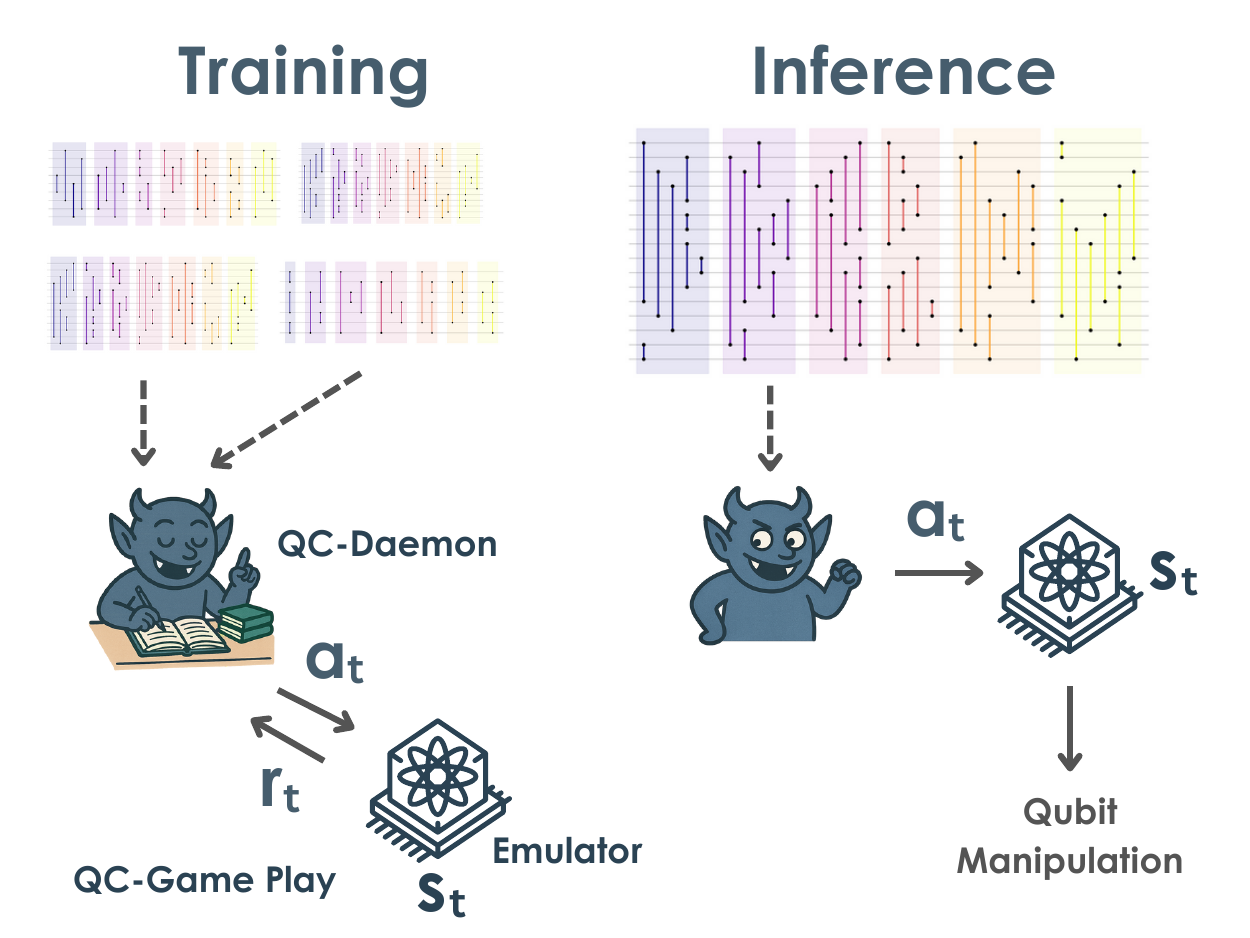}
    \caption{Training and inference with the QC-Game. During training, QC-Daemon accumulates experiences from playing the QC-Game with quantum circuits, gets feedback $r_t$, and updates its parameters. During inference, it plays the QC-Game to generate qubit manipulations to execute the target circuit based on its experiences.}
    \label{fig:qcgame}
\end{figure*}

Automated planning, often discussed in the context of artificial intelligence, robotics, and operations research, has garnered significant interest due to its ability to produce efficient plans or policies that achieve predefined objectives~\cite{ghallab2004automated,pflug2014multistage,aeronautiques1998pddl,mcdermott2003formal}. This approach is used to determine a sequence of actions---often constrained by available resources, temporal relationships, or environmental conditions---that leads from an initial state to a desired goal state. Techniques from reinforcement learning (RL)~\cite{sutton2018reinforcement} and deep learning, particularly deep RL~\cite{franccois2018introduction,silver2016mastering}, play a key role in automated planning.

The aim of this paper is to introduce automated planning of quantum device operations for executing quantum circuits.
This planning, whose goal is to maximize the fidelity---i.e., the overlap between the ideal and actual output states—of an executed quantum circuit, is crucial for the efficient quantum circuit execution.
During the execution of a circuit, the state of a device $s_t$ changes as time evolves. This state can be the physical arrangement of qubits, the error rates for each qubit, the temperature of the quantum processor, or any other classical variable we can control. 
 Inspired by the gaming testbeds for RL algorithms \cite{mnih2015human}, we call the task of finding the optimal sequence of actions that maximize the fidelity of a quantum circuit by controlling the variables in $s_t$ the \textit{quantum circuit game} (QC-Game). We formalize such planning game as a Markov decision process, and propose an AI solver that we call the quantum circuit daemon (QC-Daemon).

The recent development of quantum devices highlights the need to effectively plan and synthesize the execution of quantum device operations. For example, quantum hardware based on reconfigurable atom arrays (RAA)~\cite{levine2019parallel, bluvstein2024logical} has the freedom to change the layout of atoms, each storing a qubit, during computation. Hence, we can consider $s_t$ to be the layout and plan its transition accordingly. The recent advances of logical quantum processors~\cite{evered2023high, bluvstein2024logical,google2024,Berthusen_2024, lacroix2024scaling, aghaee2025scaling} also emphasizes the importance of careful planning of resource preparation for fault-tolerant quantum computation.

We focus on the application of automated planning to RAA, with a particular emphasis on the important role of RL in enhancing planning capabilities. To this end, we build instances of both the QC-Game and QC-Daemon tailored to this type of quantum processor. We design a QC-Game environment, termed the \textit{Atom Game}, where the QC-Daemon learns to dynamically reconfigure atom layouts while executing quantum gates. Our implementation employs a Transformer-based model having a physically motivated architecture for the QC-Daemon, enabling it to flexibly handle varying circuit structures and atom configurations. Reinforcement learning plays a crucial role in maximizing the benefits of automated planning: the RL-trained QC-Daemon achieves a considerable reduction in the total cost represented as the logarithmic infidelity across benchmark experiments involving circuits with up to 100 qubits. Importantly, we demonstrate the transferability of our approach by showing that a Transformer-based QC-Daemon trained on a diverse set of circuits can successfully apply its learned strategy to previously unseen circuits.

This paper contributes to the growing subfield of \textit{RAA move synthesis}, which maps abstract circuit executions to hardware-level instructions of RAAs. Due to their inherent flexibility, the mapping generalizes the layout synthesis problem \cite{bochen2021} on a superconducting architecture. The novel contribution of this paper is to consider the move synthesis problem from an RL perspective \cite{Pozzi2022}. Our work also contributes to the  field of AI for quantum computing \cite{alexeev2024artificial} by proposing a physically motivated architecture suitable for quantum computation. 

It should be noted that even though we focus on optimizing the reconfiguration step to simplify the problem structure for the first AI-based neutral atom compiler, the other processes may become more important depending on the problem. For example, if the circuit has a sparse and repetitive structure, optimizing the initial layout may be more important than the reconfiguration step. Our RL compiler can be extended to handle other operations, including initialization, which we leave for future work.

The rest of the paper is organized as follows. In Section~\ref{section:QC-Game}, we introduce the QC-Game and QC-Daemon. In Section~\ref{section:atom_game}, we describe the Atom Game, an instance of the QC-Game for neutral atom atom arrays. We develop a QC-Daemon to solve the Atom Game~in Section~\ref{section:solver}. Section~\ref{section:experiment} is dedicated to study the effects of the dynamic layout reconfiguration using the proposed solvers. We conclude with some discussions in Section~\ref{section:conclusion}.

\section{ The Game of Quantum Circuits}
\label{section:QC-Game}
In this section, we describe a general framework for building the QC-Daemon, an automated planning agent for quantum circuits execution. First, we design the QC-Game for optimizing the sequence of device operations in Section~\ref{section:QC-Game-mdp} and then introduce its player, the QC-Daemon, in Section~\ref{section:QC-Daemon}. 

\subsection{The QC-Game}
\label{section:QC-Game-mdp}
The QC-Game consists of finding the optimal sequence of device operations to execute a given quantum circuit while controlling the variables that affect its execution. This game can be formulated with a Markov decision processes $(\mathcal{S}, \mathcal{A}, P, R)$ as we explain below. 

Given some circuit $\mathcal{C}$, a pre-processing step decomposes the circuit into a sequence of parallelizable sets of one-qubit gates, and parallelizable sets of two-qubit gates \cite{shi2019}. This decomposition can be done in many ways, such as an ASAP heuristic \cite{tan2024compilation}. Then, we ignore all one-qubit gates for purposes of move synthesis since they can be performed locally and with much higher fidelity compared to the two-qubit gates. The resulting data is $T$ disjoint chunks representing parallel two qubit gate executions $\mathcal{C} = \{C_t\}_{t=0}^{T-1}$, where all gates in $C_t$ commute and act on independent qubits. Consequently, each chunk can be implemented in one time step. This work presumes $\mathcal{C}$ is fixed; future work may co-optimize the circuit parallelism consecutively with the move synthesis step.

Due to hardware constraints, implementing $C_t$ requires many actions, such as moving atoms around in the array to obey gate adjacency constraints.
To this end, the player of the game then plans actions for each chunk $C_t$ given a device configuration $s_t \in \mathcal{S}$ and all future chunks $\mathcal{C}_{t:T}:=\{C_j\}_{j=t}^{T-1}$. Each action $a_t \in \mathcal{A}_t$ is chosen from the set of feasible device operations at time $t$. Each action space $\mathcal{A}_t$ is a subset of $\mathcal{A}$, the set of all possible device operations. The game then evolves as $s_{t+1} \sim P(s_t, a_t, C_t)$, where $P: \mathcal{S} \times \mathcal{A} \times \mathcal{C} \longrightarrow \mathcal{P}(\mathcal{S})$ is the transition kernel---which depends on the underlying quantum device, and ends when $t$ becomes $T$. For simplicity, we let $P$ be a deterministic function $P: \mathcal{S} \times \mathcal{A} \times \mathcal{C} \longrightarrow \mathcal{S}$ in our simulations, but in real settings, this assumption might not hold.

The goal of the QC-Game is to maximize the fidelity of implementing the target circuit. However, measuring this fidelity is inconvenient in practice and thus requires a proxy that is easy to compute and relates to the circuit fidelity. Examples of such proxy include the execution time or the gate depth of the circuit, since both tend to increase decoherence and decrease circuit fidelity. We can then define a reward model $R: \mathcal{S} \times \mathcal{S} \times \mathcal{C} \longrightarrow \mathbb{R}$, based on the logarithmic infidelity, which counts actions weighted by the (log) fidelity of that action as a cost. Each action $a_t$ thus has a signal $r_t = R(s_{t}, s_{t+1}, C_t)$. This allows us to rewrite the objective of QC-Game as finding the optimal sequence of actions that maximize the cumulative reward during the circuit execution
\begin{equation}
    \left(a_t\right)^*_{t\geq 0} = \operatorname*{argmax}_{\forall t:\ a_t} \sum_{t\geq0} \gamma^{t} r_t,
\end{equation}
where $\gamma \in (0, 1]$ is a discount factor.

We emphasize that the QC-Game and what is commonly referred by \emph{quantum circuit compilation} are different and achieve complementary goals. Given a target circuit $\mathcal{C}$, quantum circuit compilation aims to reduce the depth of such circuit by creating a shorter but equivalent quantum circuit at the software level~\cite{barenco1995elementary, harrow2002efficient, dawson2005solovay}, and multiple RL algorithms have been proposed for such problem~\cite{moro2021quantum, fosel2021quantum, kremer2024practical}. In contrast, the QC-Game aims to implement $\mathcal{C}$ in the best possible way for a specific quantum processor to implement a given circuit. The QC-Game is closely related to \emph{quantum layout synthesis} and both routines share the same goal. However,  most layout synthesis algorithms only optimize device operations constrained to $C_j$ for every $j$~\cite{tan2020optimality}---similar to an iterative greedy optimization
\begin{equation}
    \left(a^*_t\right)_{t\geq 0} = \left(\operatorname*{argmax}_{a_t} r_t \right)_{t \geq 0},
\end{equation}
whereas the QC-Game does so constrained to $\mathcal{C}_{j:T}$---leveraging global information.

\subsection{The QC-Daemon}
\label{section:QC-Daemon}
Given the exponentially large state and action space of the QC-Game, we need to search an approximate solution. We focus on designing a policy function $\pi$, a well-studied technique in optimal control theory, which provides the probability distribution of actions conditioned to a given state. 

For the QC-Game, we define the QC-Daemon to be a policy function
\begin{equation}
	\pi(a_t|s_t,t, \mathcal{C}_{t:T})
\end{equation}
that is aware of the current time step $t$, the remaining quantum circuit $\mathcal{C}_{t:T}$, and the device state $s_t$. Formally, in RL, both $t$ and $\mathcal{C}_{t:T}$ are included in the definition of the state $s_t$, but we explicitly separate them to emphasize this look-ahead characteristic of the QC-Daemon.

The QC-Daemon can be implemented by using a neural network $\pi_\theta$, and in that case, each input is often converted to an embedding---a vector of $k$ real components that, after training, aims to be a better representation of the input. We note that the function that creates an embedding of the upcoming circuit $\mathcal{C}_{t:T}$, which we call a \emph{quantum circuit feature}, is reusable in different types of QC-Games. Studying the transferability of quantum circuit features across multiple QC-Game settings is an interesting direction for future research.

QC-Daemon uses QC-Game in two ways (Fig.~\ref{fig:qcgame}): Training and inference with the QC-Game. During training, QC-Daemon accumulates experiences from playing the QC-Game with quantum circuits, gets feedback $r_t$, and updates its parameters. During inference, it plays the QC-Game to generate qubit manipulations to execute the target circuit based on its experiences.

\section{Applications in reconfigurable atom arrays}
\label{section:atom_game}

In this section, we introduce the Atom Game, a specific instance of the QC-Game, designed for RAAs. The Atom Game addresses the problem of planning a sequence of layout reconfigurations, i.e. scheduling the 2D positions of each atom, for an input circuit (cf. Fig.~\ref{fig:geometry}). First, Section~\ref{section:device-overview} provides an overview of neutral atom arrays and Section~\ref{section:game-setting} explains in detail the Atom Game. 
Then, Section~\ref{section:cost} describes one proxy for circuit fidelity based on the noise introduced when moving atoms and executing gates. This proxy is closely related to the logarithmic infidelity of the circuit. Finally, Section~\ref{section:logical} briefly discusses an alternative application of the Atom Game to a neutral atom logical processor.

\subsection{Reconfigurable neutral atom arrays}
\label{section:device-overview}
Reconfigurable atom arrays (RAAs) are a quantum computing architecture based on optically trapped neutral atoms, where each atom hosts a qubit in its electronic ground states~\cite{bluvstein2022quantum}. 
Each neutral atom's position is fixed in optical traps using a spatial light modulator (SLM) laser. For simplicity, we define the two-dimensional positions of the traps as $\mathcal{V} \subset \mathbb{N} \times \mathbb{N}$. The device state is then the joint position of all $N$ atoms
\begin{equation}
s_t := \left( \textbf{v}^{(t)}_q \right)_{q=1}^{N}\in \mathcal{V}^{N},
\end{equation}
where $\textbf{v}^{(t)}_q := (c^{(t)}_q, r^{(t)}_q) \in \mathcal{V}$ denotes the position of the $q$-th atom at time $t$ and $N$ is the number of atoms. 

The movement of atoms is implemented by means of dynamical tweezers created by a laser controlled by crossed acousto-optic deflectors (AOD). The AOD laser projects both horizontal and vertical beams, forming a 2D grid of optical traps, with one trap at each intersection of every row and column. These traps can capture, move, and release atoms by turning on and off and moving the positions of the rows and columns, enabling changes of the positions $s_t$.

Single-qubit operations are performed within each occupied trap by a Raman laser system \cite{levine2022dispersive, vandersypen2005nmr}. In contrast, two-qubit operations involve bringing two atoms into close proximity and exciting them into Rydberg states with a global Rydberg laser \cite{levine2019parallel, evered2023high}. In this work, we assume a ``zoned" register layout with a storage and a gate region, as shown in Fig.~\ref{fig:geometry}.
For each two-qubit parallel gate chunk $C_t \in \mathcal{C}$, atoms are moved from a storage region to the gate region using the dynamic tweezers. Then, a laser enacts a CZ gate between each pair of atoms in the gate region. Then, atoms are moved back to the storage region in preparation for the next set of gates.

\subsection{The Atom Game}
\label{section:game-setting}

\begin{figure}
    \includegraphics[width=\linewidth]
    {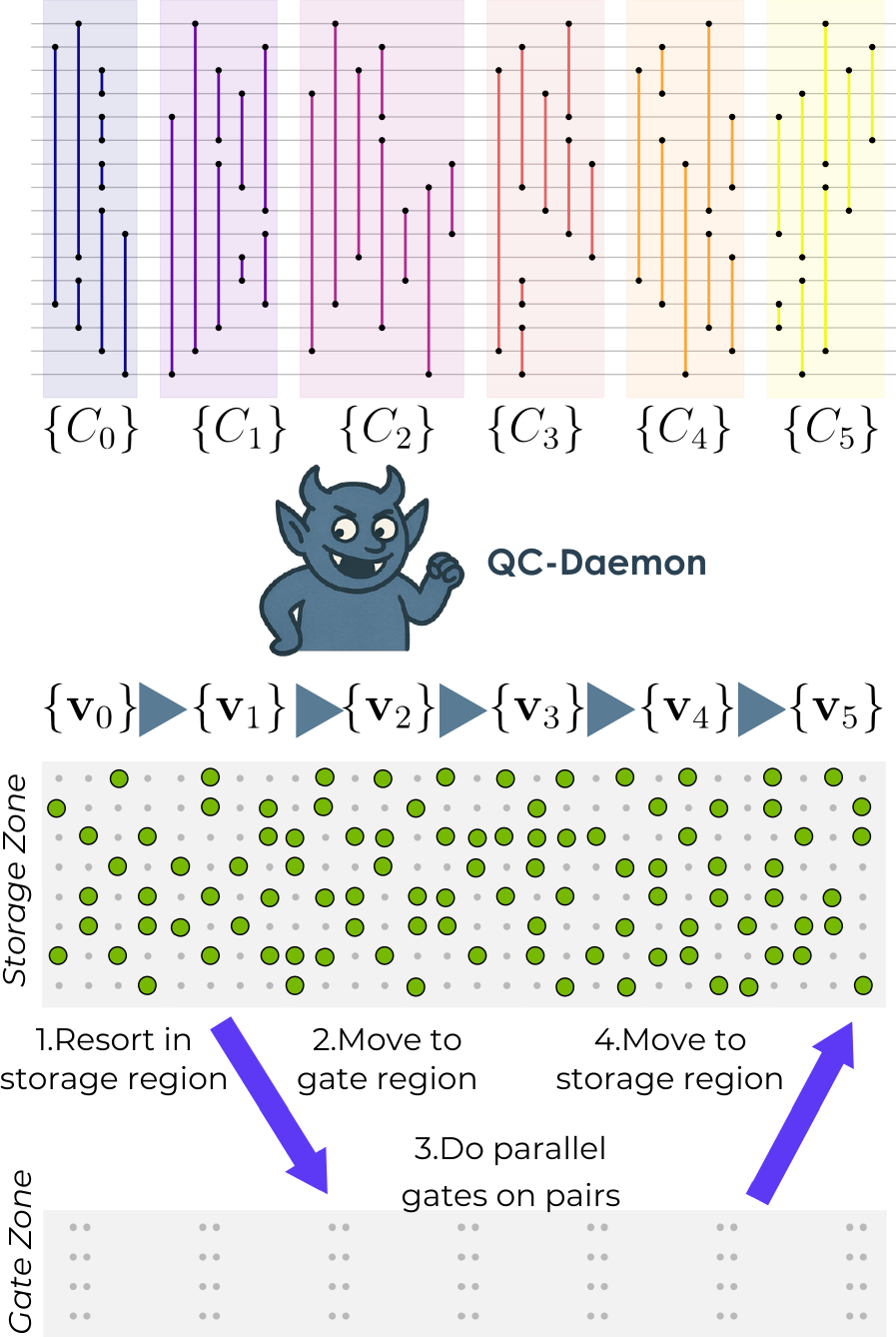}
    \caption{The structure of the atom game on a zoned reconfigurable atom array. Given a sequence of parallel 2-qubit gate chunks $\mathcal{C}=\{C_t\}_{t\geq0}$, the QC-Daemon synthesizes moves by defining atom positions $s_t$ that minimize some objective given by the efficiency of implementing that reconfiguration. Then, a back-end compiler synthesizes those reconfigurations into atom moves with the AODs. The computation is executed in 4 steps: first, atoms are resorted in the storage region; then, atoms are moved to the gate region; then a parallel entangling CZ gate is done on all pairs; then, all atoms are moved back to the same positions, and the process repeats for all chunks.}
    \label{fig:geometry}
\end{figure}

The Atom Game is a QC-Game where the goal is to choose the device state $s_t$, for each parallel gate chunk $C_t$, that minimizes the overhead cost of gate execution caused by atom reconfiguration. Having determined a state for each gate chunk, the complete circuit is executed by repeating four steps each time-step $t$, as represented in Fig.~\ref{fig:geometry}. First, atoms are loaded if the storage region is empty, or reconfigured otherwise, to state $s_t$ and then all necessary one-qubit gates are performed in this region. Second, atoms participating in the parallel two-qubit gates of the corresponding time-step are moved adjacent to each other in the gate region. Third, a parallel CZ entangling gate is done on all pairs present in the gate region. Finally, all atoms are moved back to their original locations in the storage region. This process is then repeated until the circuit is fully executed.

The problem of finding near-optimal laser control steps for a given layout of atoms $s_t$ and two-qubit gates $C_t$ has been studied in the previous literature \cite{tan2020optimal, tan2022raa, tan2024compilation, ruan2024power, ludmir2024parallax, lin2025reuse, Stade_2024,stade2025routing}. In most of these studies, the layouts before and after gate execution (i.e., the atom positions from steps 1 and 4 in Fig. \ref{fig:geometry}) within a chunk remain the same. However, Ref.~\cite{lin2025reuse} introduces a method that modifies the layout after the gate execution to minimize the atoms' moving distance to implement $C_t$.
For ease of discussion, we assume the former case of having the same layout, but our approach can be easily extended to the latter case.

In most of these previous methods, we can calculate the cost, such as the circuit infidelity, associated with applying gates $C_t$ in a given configuration of atoms $s_j$. We call the cost of such execution
\begin{equation}
    G(s_t, C_t).
\end{equation}
Similarly, for a given layout change $a_t:s_t \mapsto s_{t+1}$, we call its cost of movement
\begin{equation}
	L(s_t, s_{t+1}),
\end{equation}
and refer to $L$ as the layout function. With these cost functions, we define the reward at time $t$ as
\begin{equation}
\begin{aligned}
R(s_t, s_{t+1}, C_t) := - L(s_{t}, s_{t+1}) 
- G(&s_{t+1}, C_t)\\ 
&+G(s_0, C_t),
\end{aligned}
\end{equation}
where the last term is the gate cost when using the initial configuration $s_0$ and acts as a baseline reference.

Finally, to synthesize the actual reconfiguration actions, a lower-level compiler then converts each reconfiguration step into a sequence of valid AOD moves: first, perform $a_t$ to reconfigure the positions of atoms in the storage region, and then move atoms to execute the two-qubit gates. The process of Atom Game is summarized in Algorithm~\ref{algorithm:atomgame}. 

\begin{algorithm}[H]
\caption{One run of the Atom Game}
\label{algorithm:atomgame}
\begin{algorithmic}[1]
\Require $s_0$, $\{C_t\}_{t=0}^{T-1}$, $\pi$
\For{$t \gets 1$ to $T$}
    \State Sample $a_t \sim \pi(a_t|s_{t}, t, C_{t:T})$ 
    \State Set $s_{t+1} = P(s_{t}, a_t, C_t)$
    \State $r_{t} = - L(s_{t}, s_{t+1}) - G(s_{t+1}, C_t) +  G(s_{0}, C_t) $
\EndFor
\State \Return $\{s_t, a_t, r_t \}_{t\geq 0}$
\end{algorithmic}
\end{algorithm}

Previous works have focused on minimizing $G(s, C_t)$. However, our study explores how to reconfigure the atoms for a given function $G$ using the formalism of the QC-Game. Therefore, although we provide a specific protocol for estimating $G$ and $L$ in Section~\ref{section:cost}, our approach applies to any definition.

\subsection{Reconfiguration cost estimation}
\label{section:cost}

To reconfigure the storage region from layout $s_t$ to $s_{t+1}$ or implement a set of gates $C_t$ from layout $s_t$, the dynamic AODs must implement a sequence of moves satisfying constraints. For each chunk $C_t$, we can associate a cost---or, equivalently, a reward---that reflects the log infidelity of executing the underlying AOD moves: 
\begin{equation}
\label{eq:physical-cost}
    \mathcal{J}(D, M) = \alpha DN + \beta M.
\end{equation}
The first term represents idling error, where $D$ is the duration of the moves, $N$ is the total number of qubits in the array, and $\alpha\geq 0$ is the inverse coherence time of each qubit. The second term represents the move error, where $M$ is the number of qubits ``touched", e.g., participating in that move, and $\beta\geq0$ is a fixed cost corresponding to the fidelity loss from picking up and dropping off an atom to and from an AOD. 
By using the function $\mathcal{J}$, we model the cost for gate execution and the layout change as follows: 
\begin{align}
    L(s_t, s_{t+1}) &=\mathcal{J}\left(
        D_L(s_t, s_{t+1}), M_L(s_t, s_{t+1}) 
    \right), \\ 
    G(s_t, C_t) &= \mathcal{J}\left(
        D_G(s_t, C_t), M_G(s_t, C_t)  
    \right),
\end{align}
where $D_L(s_t, s_{t+1})$ and $M_L(s_t, s_{t+1})$ are the duration and the number of touches for the layout change $s_t \mapsto s_{t+1}$, while $D_{G}(s_t, C_t)$ and $M_{G}(s_t, C_t)$ correspond to these values when executing the gates $C_t$ from the layout $s_t$.

Instead of explicitly constructing a sequence of AOD moves to calculate a reconfiguration cost---as done in previous works \cite{tan2020optimal,tan2022raa,tan2024compilation,ruan2024power,ludmir2024parallax,lin2025reuse,Stade_2024,stade2025routing}---we instead estimate $D_L, M_L, D_G$, and $M_G$
with an upper bound that depends on the number of moves, a value obtained with a graph-theoretic heuristic. The steps to estimate $D_L (s_t, s_{t+1})$ and $M_L(s_t, s_{t+1})$ are as follows:
\begin{enumerate}
    \item \textbf{Accumulate the active participants}\\Given a layout change $s_t \mapsto s_{t+1}$, identify all atoms $q$ for which $\textbf{v}^{(t)}_q\neq \textbf{v}^{(t+1)}_q$. From that subset, identify all ``active" columns and rows 
    $$A_c = \bigcup_q \{c_q^{(t)}, c_q^{(t+1)}\}, \quad A_r =\bigcup _q\{r_q^{(t)}, r_q^{(t+1)}\}$$ in that rearrangement. The active participants $A$ are the atoms that are in one of the active rows or columns and are consistent with the crossed AOD constraints. This set may include atoms which do not change location, as they get ``caught" in the crossed AOD. 
    \item \textbf{Estimate the number of moves}
    \\Given the active participants $A$, the number of moves $n_\text{m}^{(t)}$ is computed based on a log-depth reconfiguration heuristic~\cite{Xu2024} with infinite swap space. For more details on estimating $n_\text{m}^{(t)}$, see Sec.~\ref{sec:conflictgraph2}.
    \item \textbf{Compute the total reconfiguration cost}
    \\ The total number of touches is equal to the number of moves times the number of active participants $M_L=\epsilon n_\text{m}^{(t)} |A|$ scaled by some constant $\epsilon\sim 1$ that estimates how many atoms are touched per move. Here, we assume that half the atoms are moved per step $\epsilon=0.5$. The total duration is equal to the number of moves times the characteristic time per move $D_L=n_\text{m}^{(t)}
    \uptau$. Assuming constant acceleration moves~\cite{bluvstein2024logical}, the characteristic time is the square root of the maximum distance between the start and end of all the active participants, $$\uptau=\max_{q \in A}\big[\sqrt{|\textbf{v}^{(t)}_q - \textbf{v}_q^{(t+1)}|/\gamma}\big],$$
    where $\gamma$ is a constant having the dimension of the acceleration.
\end{enumerate}

We approximate $D_G(s_t, C_t)$ and $M_G(s_t, C_t)$ with a two-step process. The first step moves the relevant pairs of atoms within the storage region in $C_t$ to be adjacent. The second step then moves such pairs of atoms to the gate region in parallel to execute $C_t$. This results in
\begin{align}
    D_G(s_t, C_t) &= D_L(s_t, \tilde{s}_t) + T_G, \\
    M_G(s_t, C_t) &= M_L(s_t, \tilde{s}_t),
\end{align}
where $\tilde{s}_t$ is the intermediate state in which atoms are reconfigured to be adjacent in the storage region. The term $T_G$ denotes the total time to move atoms from the storage to the gate region and back.

\subsubsection{The conflict graph}\label{sec:conflictgraph}

\begin{figure*}
    \centering
    \includegraphics[width=0.7\linewidth]{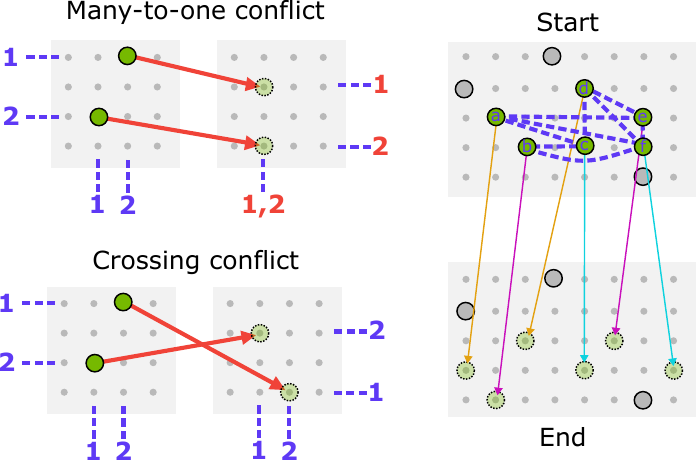}
    \caption{\textbf{Defining the conflict graph.} Left: the two kinds of parallel moves disallowed by the crossed AOD constraints. Top: a many-to-one conflict, where two AOD rows/columns (purple) merge or separate into one row/column. Bottom: a crossing conflict, where two rows/columns change order during the move. This can happen in both the X and Y directions. Right: an example reconfiguration, with active participants in green and inactive participants in grey. The conflict graph edges are shown as purple dashed; it is three-colorable, as shown in the three colours of parallel directed moves.}
    \label{fig:conflict_graph}
\end{figure*}

The \emph{Conflict Graph} $\mathcal G_c$ is a tool to schedule moves under the crossed AOD constraints and is defined as follows. Given a layout change 
$s_t \mapsto s_{t+1}$, each vertex is an atom participating in the change, with an edge between vertices if the two moves cannot be done in parallel due to violating the crossed AOD constraints~(cf Fig.~\ref{fig:conflict_graph}). There are two classes of edges in the conflict graph corresponding to the following two constraints:

\begin{itemize}
\item \textbf{Many-to-one constraint}: An AOD row cannot split into multiple rows, and multiple rows cannot merge into one. For example, in one time step
\begin{align*}
\left((2,0),\;(1,2)\right) &\mapsto \left((6,0),\;(5,1)\right) \\
\left((2,0),\;(1,2)\right) &\not\mapsto \left((6,4),\;(5,1)\right).
\end{align*}
The second move is invalid, as the AOD row picking up row~$0$ cannot split to drop onto rows~$2$ and~$3$ (Fig.~\ref{fig:conflict_graph} top left).

\item \textbf{Ordering constraint}: Two rows or columns of the AOD may not cross each other during the move, so two atoms may not move out of order. For example, in one time step
\begin{align*}
\left((2,0),\;(1,2)\right) &\mapsto \left((5,1),\;(4,4)\right) \\
\left((2,0),\;(1,2)\right) &\not\mapsto \left((5,1),\;(5,4)\right).
\end{align*}
The second move is invalid, as the AOD columns have crossed during the transfer (Fig.~\ref{fig:conflict_graph} bottom left).

Code to compute $D$ is provided here \cite{github_repo}.

\end{itemize}

\subsubsection{Estimating the number of moves $n_\mathrm{m}^{(t)}$}\label{sec:conflictgraph2}

A simple parallel move scheduler may be implemented using a vertex coloring of the conflict graph $\mathcal G_c$. A vertex colouring partitions $k$ subsets of vertices such that every partition is an independent set with no two vertices sharing an edge. If a subset of vertices on the conflict is an independent set, it can be done in a single AOD move, as no two pairs of moves conflict.

Thus, moves can be scheduled in any order by moving each partition in a single move from start to end. The number of moves equals $k$, and each atom is touched exactly once. Finding the minimum number of partitions~$k_\text{min}$, known as the chromatic number, is an archetypal NP-complete optimization problem. Brooks' theorem states that the chromatic number of a graph is upper bounded by the maximum connectivity of the graph~\cite{Brooks1941}, which can be found efficiently with a greedy heuristic. Thus, an (over)estimate of $n_\mathrm{m}^{(t)}=\Delta(\mathcal G_c)+1$, the maximum connectivity of the conflict graph.

A more advanced move scheduler can be implemented based on a divide-and-conquer approach first introduced in \cite{Xu2024} by adding intermediate moves in a swap region to some intermediate positions $s_t \to s^{\prime}\to s^{\prime\prime}\to\cdots\to s_{t+1}$. Crucially, {the new conflict graph of each intermediate move is the union of the two induced subgraphs of the original conflict graph}, plus extra edges from the many-to-one constraint. If an extensive number of edges can always be removed in this manner with a MaxCut heuristic, the number of moves required to remove all edges and thus reconfigure the layout is logarithmic in the number of edges in the conflict graph $n_\mathrm{m}^{(t)}=\delta \log\big(|V(\mathcal G_c)|\big)$. This is a divide-and-conquer approach, as the conflict graph is iteratively partitioned into smaller and smaller graphs until the resorting is complete. It is beyond the scope of this work to analyze the specific action of this iterative partitioning. Here we simply assume that the swap region is large enough so that the landscape is convex and a greedy MaxCut heuristic always succeeds. The particular exponent of the partitioning is represented by $\delta\sim 1$ which we assume to be order 1 in this work.

Thus, to estimate the number of moves, we count the number of edges in the conflict graph and take the log (in any base). The iteration exponent $\delta$ is then an overall scale factor of the cost function $n_\mathrm{m}^{{(t)}}\to \delta n_\mathrm{m}^{(t)}$, which is irrelevant to the action of the QC-Daemon.

\begin{figure*}
    \centering
    \includegraphics[width=1\linewidth]{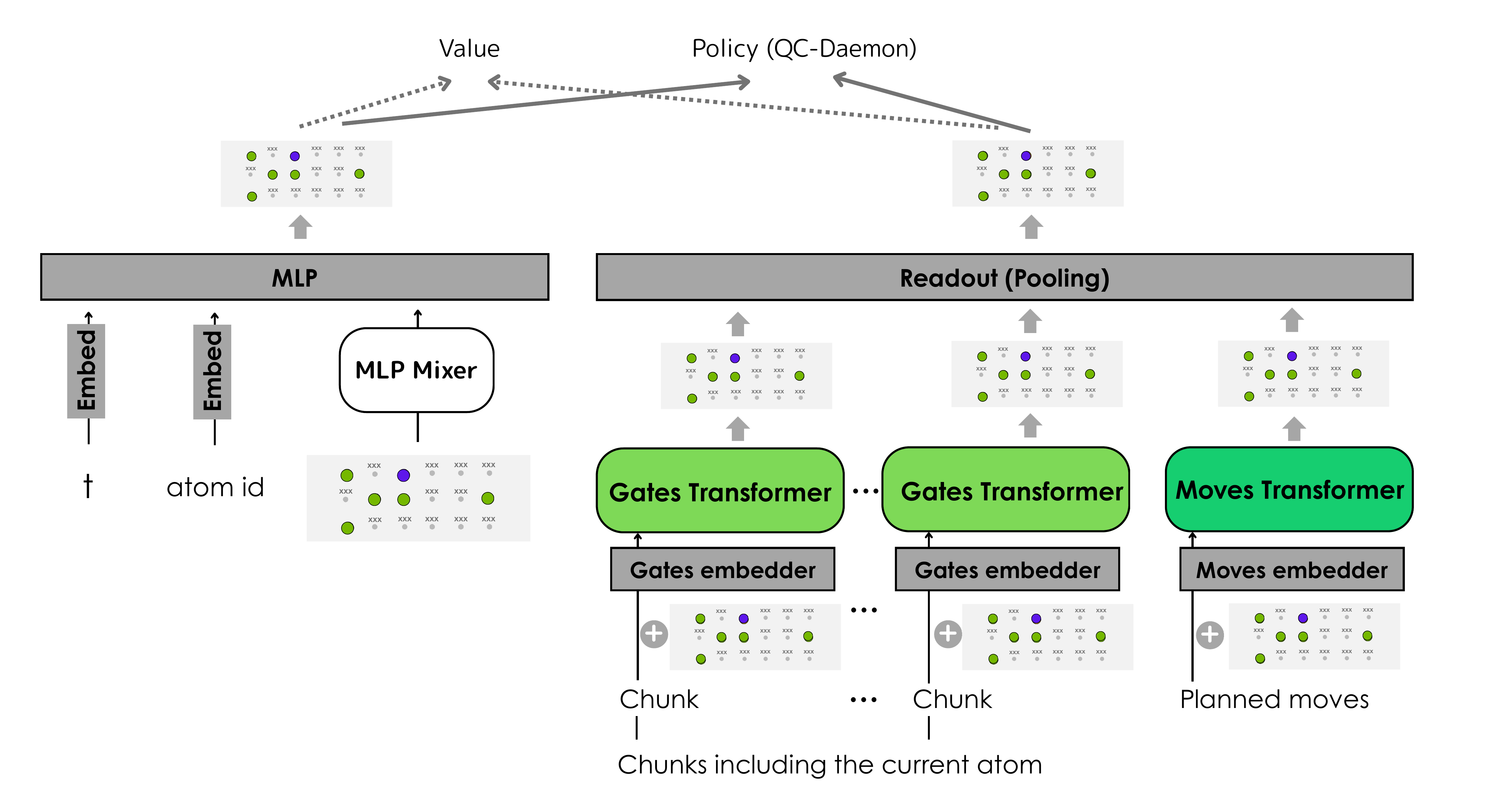}
    \caption{The implementation of the atom-by-atom policy (QC-Daemon) $\pi^A$ and the value function. Our model comprises two parts: static feature extraction (left) and dynamic feature extraction, which are implemented using the Gate Transformer and Move Transformer (right). The static features are independent of the circuit and the planned moves, and indicate which traps are more favorable for each atom at each time step. The dynamic features are conditioned on the circuit and the already planned moves.}
    \label{fig:network}
\end{figure*}

\subsection{Remark on logical processor}
\label{section:logical}
Even though we demonstrate the framework for moving individual physical qubits, we can also use the same Atom Game framework for moving groups of physical qubits in a coarse-grained manner. This opens the door to actively optimizing the layout of the qubits in an error-corrected quantum computer using reinforcement learning, as the constraints of such a system may be similar to those present in crossed AOD grids~\cite{Xu2024}. 

Using our compilation approach on groups of atoms, where each group corresponds to a particular logical qubit and is moved all together, can enable larger amounts of parallelization in the logical transversal gates of a circuit~\cite{bluvstein2024logical,rodriguez2024}. Furthermore, abstracting away the hardware implementation, the Atom Game framework could be used for circuit compilation in topological codes, where logical qubits are ``carved out" on a grid, as it happens with surface codes~\cite{kitaev2003fault, nayak2008non}. In such cases, logical qubits can be defined via holes or patches in a grid, and logical operations are performed via braiding~\cite{raussendorf2007fault} or lattice surgery~\cite{horsman2012surface}, respectively. Thus, each logical qubit is defined within a spatial region in the topological code grid that changes throughout the circuit implementation~\cite{litinski2019game}, and consequently, will have a cost associated to its movement with such a grid.

Previously, surface code compilation has been solved with heuristic algorithms \cite{paler2020opensurgerytopologicalassemblies, beverland2022surface, watkins2024high}. These compilers optimize over the logical lattice instructions that allow implementing quantum algorithms, such as entangling measurements with lattice surgery merges and splits, distillation of magic states in specific regions, and other instructions that are required for universality. Given its resemblance to the RAA compilation problem, the QC-Game framework would similarly benefit circuit compilation at the logical circuit level, accounting for information encoded in the reward function that would otherwise be challenging to include in non-data-driven methods.

\section{Playing the Atom Game}
\label{section:solver}
In this section, we introduce our QC-Daemon for Atom Game. Our QC-Daemon determines the actions of each `playable atom' for each time $t$. 
The playable atoms are chosen in the following manner. Let $f$ be the function to extract the list of qubit indices included in $C_t$. The playable atoms $P_t$ are defined as
\begin{equation}
	P_t = \bigcup_{w=0}^{W-1} f(C_{t+w}),
\end{equation}
where $W$ (Window Size) is the hyperparameter. Namely, the atoms included in the gates in the next $W$ chunks are used as the playable atoms.

\subsection{Auto regressive action generation}

The action generation is performed by determining the action of each playable atom in auto-regressive manner. Let $P_t = \left(q_b\right)_{b=1}^{|P_t|}$. The QC-Daemon policy can be decomposed as 
\begin{equation}
\begin{split}
\pi(a_t|s_{t}, t, C_{t:T}) = \prod_{b} \pi^A\left(a_t^{(b)}|s_{t}, t, C_{t:T}, q_b, \left(a_t^{(b^{\prime})}\right)_{b^{\prime}=1}^{b-1}\right),
\end{split}
\end{equation}
where the function $\pi^A$ is an atom-by-atom policy that determines the action for each playable atom $q_b$. Each action specifies only the next position of the atom. The action $a_t^{(b)}$ generated for atom $b$ is used in subsequent action generations for each playable atom. These individual actions are finally aggregated into the overall action as $a_t = \left(a_t^{(b)}\right)_{b=1}^{|P_t|}$.

\subsection{Our model}
Here, we describe our model for $\pi^A$. For notational convenience, we write $s_t = \left( \textbf{v}_q \right)_{q=1}^{N}$ and 
the layout after applying $\left(a_t^{(b^{\prime})}\right)_{b^{\prime}=1}^{b-1}$ to $s_t$ as $\left( \textbf{v}_q^{\prime} \right)_{q=1}^{N}$ in this section. We also write the current playable atom as $q_0$.

Our model comprises two parts: static feature extraction and dynamic feature extraction. The static features do not depend on the circuit or the planned moves, and describe which traps are more favorable for each atom at each time step. The dynamic features are conditioned on the circuit and the already planned moves.
The input of static features is the time $t$, the playable atom's id, and $\left( \textbf{v}_q^{\prime} \right)_{q=1}^{N}$. In contrast, the input of the dynamic features encompass the subsequent chunks $\textbf{c} = C_{t_1}, \cdots, C_{t_K}$ that include the current playable atom, where $K$ (Horizon-Length) is a hyper-parameter, as well as $\left( \textbf{v}_q \right)_{q=1}^{N}$ and $\left( \textbf{v}_q^{\prime} \right)_{q=1}^{N}$.

\subsubsection{Static feature extraction}
The input of the static feature extraction is converted into a vector embedding:
$t \rightarrow \textbf{e}_t$, 
\text{atom id} $\rightarrow \textbf{e}_a$, and 
$(\textbf{v}_q^{\prime})_{q=1}^{N} \rightarrow \textbf{e}_l$.
They are then concatenated into a vector $\textbf{e}$:
\begin{equation}
\textbf{e} = \textbf{e}_t \oplus \textbf{e}_a \oplus \textbf{e}_l
\end{equation}
Finally, a multilayer perceptron is applied to $\textbf{e}$ to obtain $\textbf{d} := \{d_j\}_{j=1}^{n_{\textrm{grid}}}$. 

For building $\textbf{e}_l$, we use the MLP-Mixer \cite{tolstikhin2021mlp}. The MLP-Mixer is a neural network architecture that relies solely on multi-layer perceptrons (MLPs), without using convolutions or attention mechanisms. It operates on image patches rather than individual pixels, processing data through two types of MLPs: one mixes information across spatial (patch) dimensions and another mixes across channel dimensions. This separation allows the MLP-Mixer to capture local and global data patterns.
In our case, we use $\left( \textbf{v}_q^{\prime} \right)_{q=1}^{N}$ as the input to the MLP-Mixer, treating each spatial cell as a patch embedding that contains one-hot encoded atom assignments.

\subsubsection{Dynamic feature extraction}

Two types of Transformers, the Gate Transformer and the Move Transformer, are used to extract dynamic features. The output from those components is combined and processed in the readout layer. 
\paragraph{Gates Transformer} In each chunk, we consider the gate containing the current playable atom (PA-gate) along with other gates (non-PA-gates) included in the chunk. The Transformer layer captures correlations between these non-PA-gates and the PA-gate by assuming that the playable atom moves to each position of the grids. 

The input to the Transformer layer is constructed via a gate embedder, whose output is 
$E_{\text{atoms}}^g \oplus E_{\text{grids}}^g$, 
where $E_{\text{atoms}}^g$ includes $n_{\text{atoms}}$ embeddings and $E_{\text{grids}}^g$ includes $n_{\text{grids}}$ embeddings. 
For each position of the traps $\textbf{v}$, we define learnable embeddings $\textbf{e}_g(\textbf{v})$. 
Then the element of $E_{\text{atoms}}$ is 
\begin{equation}
\begin{split}
    &[E_{\text{atoms}}^g]_q \\
    &= \left\{
    \begin{array}{cc}
        \textbf{e}_g(\textbf{v}_q^{\prime}) - \textbf{e}_g(\textbf{v}_{q^{\prime}}^{\prime}) & \text{the non-PA gates include $q$}\\
        \textbf{0} & \text{otherwise}
    \end{array}
    \right., 
\end{split}
\end{equation}
where $q^{\prime}$ is the ID of the atom that couples with $q$-th atom in the non-PA gate. 
Let $\textbf{v}^{(1)} \cdots \textbf{v}^{(n_{\textrm{grids}})}$ be the positions of all traps. Then the element of $E_{\text{grids}}$ is 
\begin{equation}
    [E_{\text{grids}}^g]_j = \textbf{e}_g(\textbf{v}^{(j)}) - \textbf{e}_g(\textbf{v}_{q_0}^{\prime}).
\end{equation}

The Gates Transformer is composed of multiple layers of (i) the multi-head attention, (ii) the layer norm, and (iii) the multi-layer perceptron. In the multi-head attention layer, the information of each vector at each position propagates to the output of different positions; how the information propagates is governed by the so-called attention mechanism \cite{vaswani2017attention}, but the attention mask prohibits some propagation. We accept all the information flow at the same position, namely, the information at a position can be included in the output at that position. 
For the information flow to a different position, we only accept the flow from $E_{\text{atoms}}^g$ to $E_{\text{grids}}^g$. In addition, we prohibit any flow from the position where $[E_{\text{atoms}}^g]_q = \textbf{0}$. The attention mask is setup to realize the information flow and is used in all the multi-head attention layers.

The output of the Gates Transformer has the same structure as $E_{\text{atoms}}^g \oplus E_{\text{grids}}^g$, and we use the last $n_{\textrm{grids}}$ vectors in the later calculation. We write the list of vectors obtained by applying Gates Transformer to the $k$-th input chunk as $F^{(k)} := \{\textbf{f}^{(k)}_j\}_{j=1}^{n_{\textrm{grids}}}$.

\paragraph{Moves transformer} The moves already planned are processed by the Moves Transformer. In this component, the Transformer layer captures correlations between already planned moves and potential moves of the current playable atom. The input to the Transformer is built via the moves embedder, producing
$E_{\text{atoms}}^m \oplus E_{\text{grids}}^m$, 
where $E_{\text{atoms}}^m$ includes $n_{\text{atoms}}$ embeddings and $E_{\text{grids}}^m$ includes $n_{\text{grids}}$ embeddings. 
As in the case of the gates embedder, we define learnable embeddings $\textbf{e}_m(\textbf{v})$ for each grid position $\textbf{e}_m$. Then, the element of $E_{\text{atoms}}^{m}$ is 
\begin{equation}
    [E_{\text{atoms}}^{m}]_q = 
       \textbf{e}_m(\textbf{v}_q^{\prime}) - \textbf{e}_m(\textbf{v}_q) 
\end{equation}
For $E_{\text{grids}}^m$, we set $[E_{\text{grids}}^{m}]_j = \textbf{e}_m(\textbf{v}^{(j)}) - \textbf{e}_m(\textbf{v}_{q_0})$. 

The architecture of the Moves Transformer is the same as the Gates Transformer. The attention mask is also designed similarly. We accept all the information flow at the same position. For the information flow to a different position, we only accept the flow from $E_{\text{atoms}}^m$ to $E_{\text{grids}}^m$. In addition, we prohibit any flow from the position where $[E_{\text{atoms}}^{m}]_q = \textbf{0}$. 

The output of the Moves Transformer has the same structure as $E_{\text{atoms}}^m \oplus E_{\text{grids}}^m$, and we use the last $n_{\textrm{grids}}$ vectors in the later calculation. We write the list of vectors obtained by applying Moves Transformer as $F^{(0)} := \{\textbf{f}_j^{(0)}\}_{j=1}^{n_{\textrm{grids}}}$.

\paragraph{Readout} The outputs from Gates Transformer and Moves transformer is aggregated by the following mean pooling step to $F = \{\textbf{f}_j\}_{j=1}^{n_{\textrm{grids}}}$, where 
\begin{equation}
    \textbf{f}_j := \frac{1}{K + 1}\sum_{k=0}^K \textbf{f}_j^{(k)}.
\end{equation}

\subsubsection{The QC-Daemon's policy and its value function}
$F$ and $\textbf{d}$ are used to calculate the policy (QC-Daemon). We also describe how the value function is constructed for the actor-critic type RL algorithm. 

\paragraph{Policy (QC-Daemon)} For each position, MLP is applied to $\textbf{f}_j$ for each $j$, and converted to one-dimensional output $o_j$. 
The logit is defined by $w_j := o_j + d_j$. The grid ID $j$ is sampled according to the probability proportional to $\exp(w_j)$, but the grid IDs filled with other atom are masked.

\paragraph{Value} 
We first mask the grid IDs filled with other atom; let $\{j_1, \cdots, j_m\}$ be the list of IDs unmasked, then one vector $\textbf{f}$, which is the average of $\textbf{f}_{j_1} \cdots \textbf{f}_{j_m}$ is calculated. Then, MLP with one-dimensional output is applied to $\textbf{f}$ and another MLP with one-dimensional output is applied to $\textbf{d}$. Finally, these outputs are summarized as the output of the value function.

\section{Numerical Experiments}
\label{section:experiment}
\begin{figure*}
    \centering
    \includegraphics[width=0.95\linewidth]{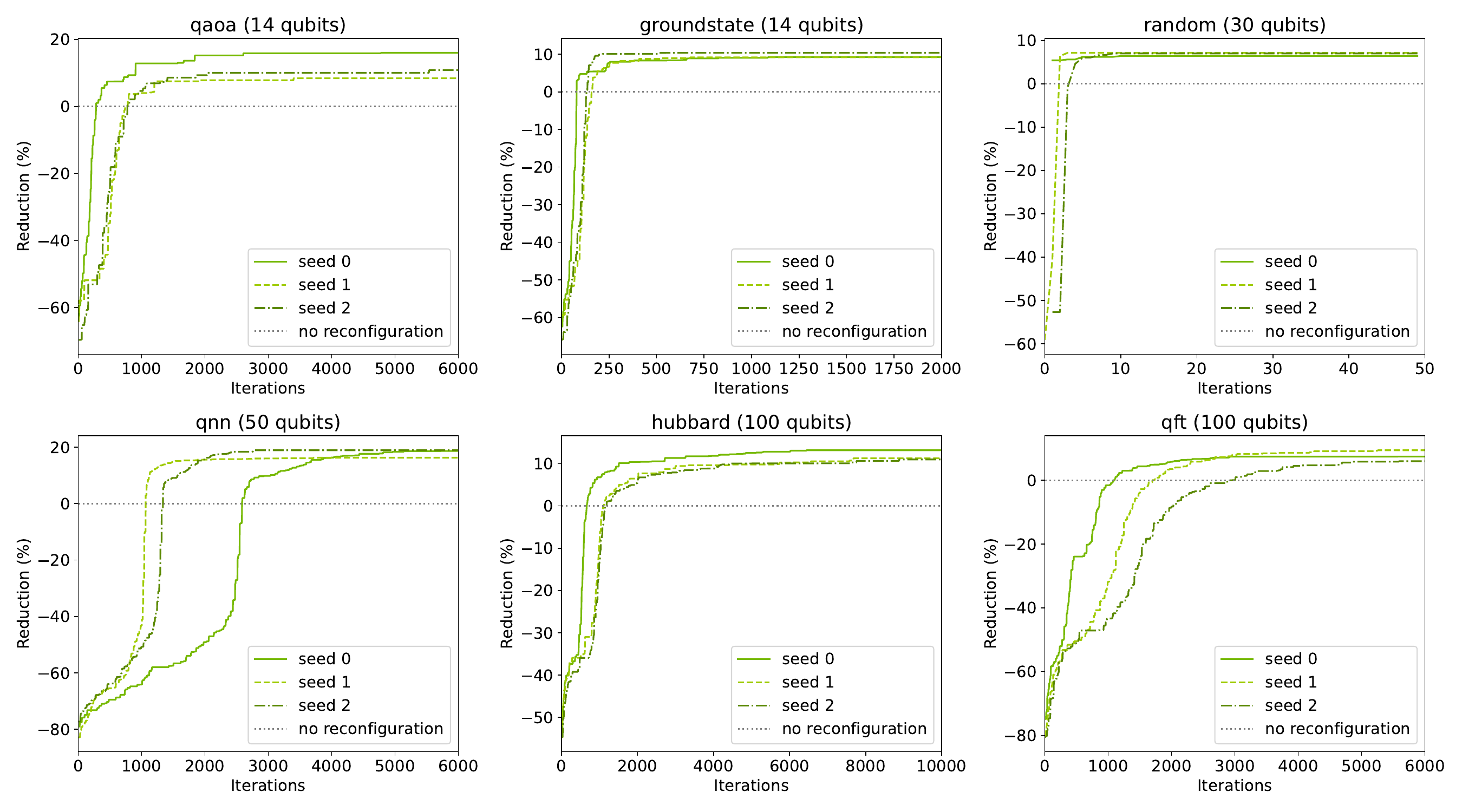}
    \caption{Cost reduction (\%) for each benchmark across training iterations, as achieved by our QC-Daemon in the experiment described in Section~\ref{section:scratch}. The model is trained from scratch and applied to the same circuit. The horizontal axis shows the number of training iterations, while the vertical axis indicates the cost reduction relative to the baseline without reconfiguration (gray dotted line). Each green line represents a training run with a different random seed for initializing the configuration and model parameters.}
    \label{fig:cost-reduction}
\end{figure*}
\begin{figure*}
    \centering
    \includegraphics[width=0.95\linewidth]{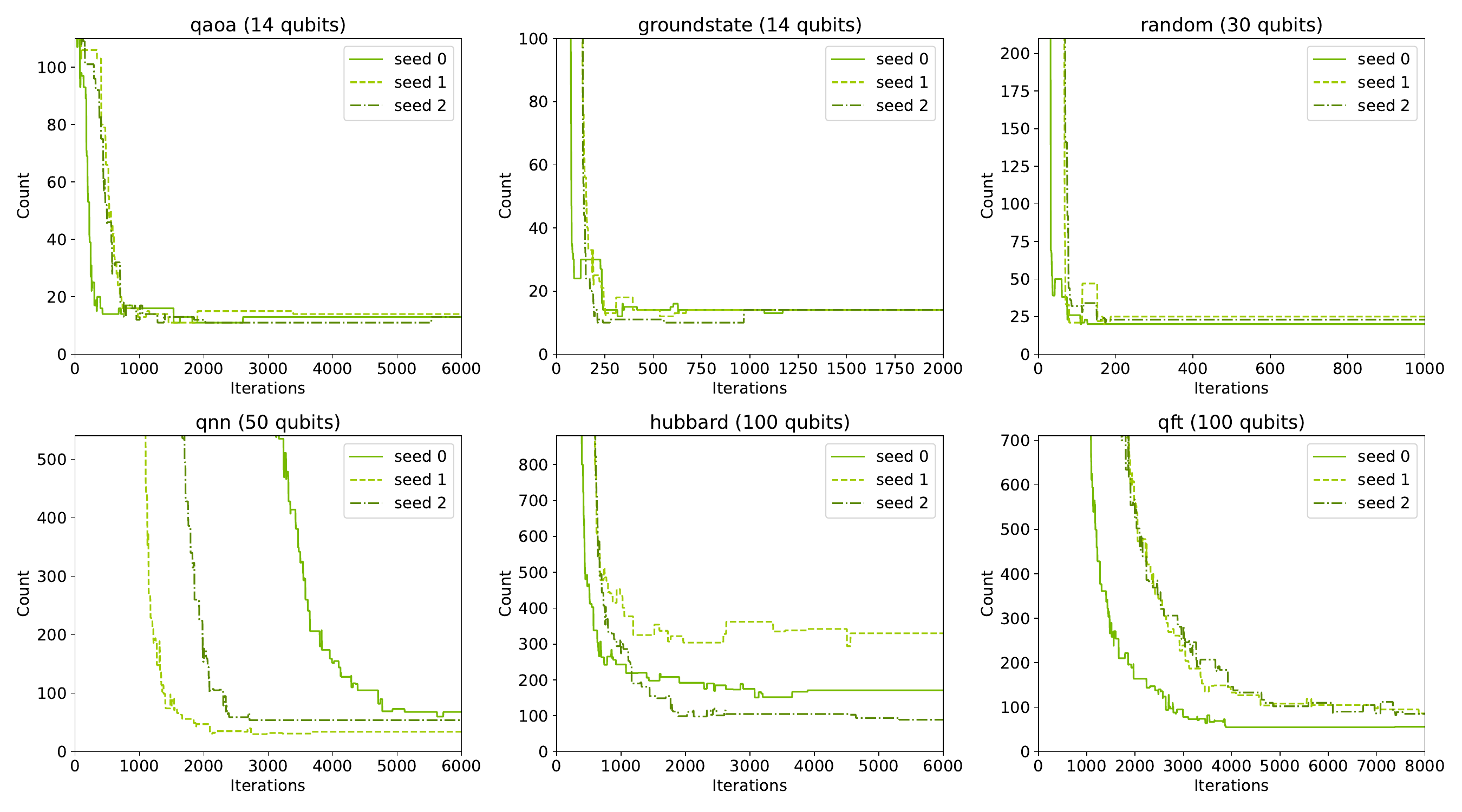}
    \caption{The number of reconfiguration actions for each benchmark at each iteration in the same experiment as Fig.~\ref{fig:cost-reduction}. We count only the actions that result in a playable atom being placed at a different grid location, thereby changing the overall layout.}
    \label{fig:layout-change}
\end{figure*}
In this section, we show two types of numerical experiments. 
In the first experiment in Section~\ref{section:scratch}, we train the model with one circuit and apply it to that same circuit. 
In the second experiment in Section~\ref{section:transferability}, we first train the model with 30 circuits with different numbers of qubits and apply it to new unseen problems to verify the transferability of our model.

\subsection{Training from scratch}
\label{section:scratch}
We use quantum circuits for quantum Fourier transform (qft), quantum neural network (qnn), variational quantum eigensolver (groundstate), quantum approximate optimization algorithm (qaoa) from the MQT Bench \cite{quetschlich2023mqtbench} and generate Hamiltonian evolution circuits for the Fermi-Hubbard model \cite{arute2020observation} (hubbard). 
For the Fermi-Hubbard model simulation, we use the first-order Lie-Trotter formula to decompose the time evolution, repeating the operation twice. Specifically, for a given time evolution $e^{iHt}$ with $H$ as the Hamiltonian having the Pauli decomposition $H = \sum_{j} c_j P_j$, we obtain the circuit for our simulation by
$$
\left (\prod_{j} e^{i c_j P_j \Delta t}\right)^{t/\Delta t}, \quad c_j \in \mathbb{R}\ \text{and} \ t/\Delta t=2.
$$

For calculating the reconfiguration cost of $L$ and $G$, we set the physical parameters in Eq. \eqref{eq:physical-cost} to $\alpha = 0.02$ and $\beta = 0.002$. We use proximal policy optimization (PPO) \cite{schulman2017proximal} as the RL algorithm. The other hyper-parameters are listed in Appendix~\ref{section:hyper-parameter}. The training is performed on a system equipped with eight NVIDIA H100 Tensor Core GPU.

\begin{figure*}
    \centering
    \includegraphics[width=1\linewidth]{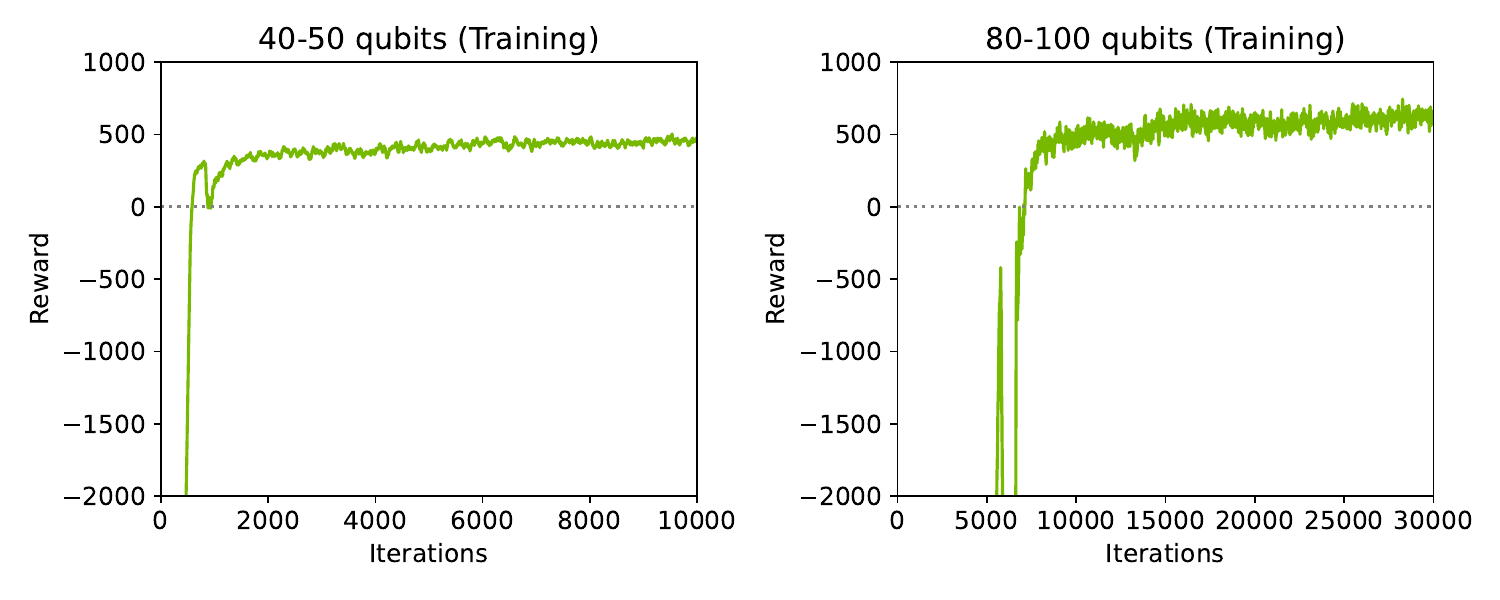}
    \caption{The change in the mean reward for each iteration during the training of the transferable models with 40–50 qubit circuit datasets (left) and 80–100 qubit circuit datasets (right). The mean reward is calculated as the moving average over the past 50 steps. A positive reward indicates that the cost is reduced compared to the case without reconfiguration.  }
    \label{fig:reward_transfer}
\end{figure*}

In Fig.~\ref{fig:cost-reduction}, we show the cost reduction ratio at each iteration for each benchmark problem. The horizontal axis represents the number of training iterations, and the vertical axis indicates the cost reduction compared to the case where no reconfiguration is performed, which corresponds to the gray dotted line. Each green line corresponds to the training with a different seed for the random generation of the initial configuration and model parameters. We see that the cost reduction is negative at the beginning of training, which means that the generated actions increase the cost due to unnecessary layout changes. However, after some iterations, the cost reduction becomes positive, and we can achieve a reduction in cost of up to about 20\% through layout changes.
In Fig.~\ref{fig:layout-change}, we show the number of actions that change the layout of atoms; namely, we count only the actions that specify a position for a playable atom different from its current grid. We see that at the beginning of training, many actions are generated, but only a few remain after training. This implies that the QC-Daemon successfully learns how to generate effective layout changes.

\subsection{Transferability evaluation}
\label{section:transferability}

We also conducted an experiment to evaluate the transferability of QC-Daemon. In this experiment, the model was built using only dynamic feature extraction, allowing it to generalize across different qubit counts and circuit structures. An attention mechanism, which enables the model to handle variable-length inputs, makes this possible. 

As the training dataset, we generated 30 Hamiltonians, each consisting of 40 terms of local random Pauli operators acting on 40–50 qubits. These were converted into quantum circuits by applying 4 steps of Trotterization, and used for training. During training, one Hamiltonian was randomly selected from the 30, and the model was executed starting from a random initial configuration. The model was then updated based on the execution result, using the same hyperparameters as in previous experiments. 

As the test dataset, we additionally generated three new test circuits not included in the training dataset with 40–50 qubits using the same procedure as described above. We compared the best outcomes from 1-shot, 10-shot, and 100-shot evaluations to those obtained without training, where $n$-shot denotes the best result obtained from $n$ independent runs of the trained model on the same circuit. 

We also repeated the same procedure using larger Hamiltonians with 60 terms acting on 80–100 qubits, and studied the performance. The training is performed on a system with six NVIDIA H100 Tensor Core GPUs.

In Fig.~\ref{fig:reward_transfer}, we show the change in the mean reward for each iteration during the training of the transferable models with 40–50 qubit circuit datasets (left) and 80–100 qubit circuit datasets (right). 
The reward is calculated as the cost reduction compared to the case without reconfiguration. A positive reward indicates that the cost is reduced compared to the case without reconfiguration. 
The mean reward is calculated as the moving average over the past 50 steps.
In both cases, the mean reward becomes positive after several thousand iterations, indicating that the model successfully learns generally effective reconfiguration strategies across multiple circuits.

Table~\ref{tab:transfer_comparison_1} and Table~\ref{tab:transfer_comparison_2} show the cost reduction for test circuits with and without transfer for the 40–50 qubit circuit datasets (left) and the 80–100 qubit circuit datasets, for each number of shots. Even in the 1-shot case, we observe that the model trained with the datasets can successfully generate practical reconfiguration steps for unseen circuits, though the untrained model generates unreasonable layout changes. Table \ref{tab:transfer_comparison_3} shows the duration (seconds) for generating all the actions for each number of shots using the same test circuit as in Table~\ref{tab:transfer_comparison_2} on a system equipped with an NVIDIA H100 Tensor Core GPU. We also show the number of chunks in the same table. We see that the difference in execution time between 1 shot and 10 shots is smaller than a factor of two due to efficient parallel computation. The execution time with 100 shots scales slightly worse from the 10-shot case, but is still better than linear scaling. We note that most of the computational time was dedicated to the action generation from the model on the GPU. If we use a reward that requires intensive CPU resources, the scaling behaviour would degrade. 

\begin{table}[ht]
\centering
\caption{Cost reduction with and without transfer in the 40–50 qubit experiments for each number of shots.}
\begin{tabular}{cccccc}
\toprule
Seed & Qubits & Transfer & 1 shot & 10 shots& 100 shots \\
\midrule
\multirow{2}{*}{0} & \multirow{2}{*}{48} & No & -74.95\% & -72.91\% & -72.31\% \\
 && Yes & 8.36\% & 8.36\% & 8.90\% \\
\hline
\multirow{2}{*}{1} & \multirow{2}{*}{42} & No & -78.42\% & -76.77\% & -73.85\% \\
 && Yes & 7.42\% & 8.52\% & 10.33\% \\
 \hline
\multirow{2}{*}{2} & \multirow{2}{*}{46} & No & -62.85\% & -62.81\% & -57.49\% \\
 && Yes & 8.32\% & 8.60\% & 9.20\% \\
\bottomrule
\end{tabular}
\label{tab:transfer_comparison_1}
\end{table}

\begin{table}[ht]
\centering
\caption{Results of the same experiment as in Table~\ref{tab:transfer_comparison_1}, but conducted with 80–100 qubit circuits for each number of shots.}
\begin{tabular}{cccccc}
\toprule
Seed & Qubits & Transfer & 1 shot & 10 shots & 100 shots \\
\midrule
\multirow{2}{*}{0} & \multirow{2}{*}{97} & No & -86.98\% & -86.43\% & -85.31\% \\
& & Yes & 6.66\% & 8.51\% & 8.51\% \\
\hline
\multirow{2}{*}{1} & \multirow{2}{*}{84} & No & -78.37\% & -78.37\% & -77.21\% \\
 && Yes & 7.90\% & 8.66\% & 8.85\% \\
\hline
\multirow{2}{*}{2} & \multirow{2}{*}{92} & No & -92.18\% & -91.64\% & -90.86\% \\
 && Yes & 7.06\% & 7.78\% & 8.57\% \\
\bottomrule
\end{tabular}
\label{tab:transfer_comparison_2}
\end{table}

\begin{table}[ht]
\centering
\caption{The duration (seconds) for generating all the actions for each number of shots using the same test circuit as in Table~\ref{tab:transfer_comparison_2} on a system equipped with an NVIDIA H100 Tensor Core GPU.}
\begin{tabular}{cccccc}
\toprule
Seed & Qubits & Chunks & 1 shot & 10 shots & 100 shots \\
\midrule
0 & 97 & 436 & 12.83  & 18.32 & 87.02 \\
1 & 84 & 490 & 12.45  & 17.44 & 81.74 \\
2 & 92 & 335 & 13.12  & 17.60 & 85.84 \\
\bottomrule
\end{tabular}
\label{tab:transfer_comparison_3}
\end{table}

The transferability shown in the experiment is practically essential, as it provides a way to generate moves without additional training. While we restrict the dataset class in this demonstration, it would be beneficial to pre-train the model on various quantum circuits as a foundation model for generating layout changes in future research.

\section{Discussions and conclusions}
\label{section:conclusion}

In this paper, we introduced the QC-Game as a framework based on Markov decision processes to dynamically control the state of a device while executing a given quantum circuit. We then propose QC-Daemon as an agent to solve the QC-Game. An essential application of the QC-Game is the Atom  Game, which involves planning the sequence of reconfiguring the layout of atoms in reconfigurable neutral atom arrays. The actual physical operations for reconfiguration and gate execution in the Atom Game are handled abstractly, making the approach compatible with any reconfiguration or gate-execution protocol. We demonstrated that our reinforcement-learning agent, QC-Daemon, constructed with two types of transformers, can reduce the logarithmic infidelity compared to the scenario where no layout change is made, across various benchmark problems. We also demonstrate the transferability of our model using datasets of 40–50 qubits and 80–100 qubits. This transferability is beneficial for rapidly generating physical operations in practical applications.

There are multiple directions for future research. As we note in Section~\ref{section:introduction}, this work focuses on optimizing the reconfiguration step to simplify the problem structure. However, other stages in the compilation pipeline may become more important depending on the nature of the target circuit such as initialization. Extending our AI compiler to support additional stages is an important direction for future research.

Our reward provides a rough estimate of the infidelity, and we do not perform a full simulation of scheduling gates or reconfigurations on an actual quantum device. Performing such simulations would require assuming specific details of a real scheduler. In principle, this would only involve modifying the definitions of $G(s_t, C_t)$ and $L(s_t, s_{t+1})$. 
Additionally, in our setting, the reconfiguration/gate execution protocols are given as inputs. However, another possibility is training the agent to produce the physical atom moves for reconfiguration or gate execution directly, given a particular atom layout. The application for fault-tolerant quantum computation (FTQC) is also an interesting direction. We expect that the same argument applies in FTQC by using a logical qubit instead of a physical one as the unit of the atom move; however, additional operations necessary for a logical processor, such as the implementation of non-transversal gates, will affect the advantage of our QC-Daemon.

\section*{Acknowledgement}
A related work \cite{stade2025routing} considering the reconfiguration process was posted almost simultaneously with this manuscript. Although developed independently, their study addresses a similar problem and provides complementary insights. We acknowledge their timely contribution to this rapidly evolving field.

The authors thank early discussions with Shengtao Wang and Casey Duckering. H.H. acknowledge support from the NSERC-Google Industrial Research Chair award. L.M.C. acknowledges support from the Novo Nordisk Foundation, NNF Quantum Computing Programme. K.P. acknowledges the generous support of the Canada 150 Research Chairs
Program through A.A.-G. A.A.-G. thanks Anders G. Frøseth for his generous support and acknowledges the generous support of Natural Resources Canada and the Canada 150 Research Chairs program. This research is part of the University of Toronto’s Acceleration Consortium, which receives funding from the CFREF-2022-00042 Canada First Research Excellence Fund.

\bibliographystyle{ieeetr}
\bibliography{main.bib}

\newpage
\begin{widetext}
\appendix

\section{Hyper-parameter of the numerical experiment}
\label{section:hyper-parameter}

\newcolumntype{P}[1]{>{\raggedright\arraybackslash}p{#1}}

\subsection{Machine learning model parameters}
\begin{table*}[htbp]
    \centering
    \renewcommand{\arraystretch}{1.2} 
    \caption{Model architecture, training, and experimental parameters. The device parameters are used to compute a dimensionless quantity, the infidelity. Therefore, while each parameter may originally have a physical unit, the units are omitted in the description, and only scalar values are shown.}
    \begin{tabular}{@{}l@{\hspace{1cm}}P{7cm}@{}} 
        \toprule
        \textbf{Category} & \textbf{Parameters} \\
        \midrule
        
        \textbf{Dynamic feature:} Gate transformer
            & Embedding dimension = 20 \\
            & Number of layers = 3 \\
            & Number of heads in multi-head attention = 4 \\
        \midrule
        
        \textbf{Dynamic feature:} Move transformer
            & Embedding dimension = 20 \\
            & Number of layers = 3 \\
            & Number of heads in multi-head attention = 4 \\
        \midrule
        
        \textbf{Static feature}
            & Player embedding dimension = 40 \\
            & Time embedding dimension = 40 \\
            & Board embedding dimension = 10 \\
            & Number of layers in MLP-Mixer = 1 \\
        \midrule
        
        \textbf{Training parameters:} PPO 
            & Clipping coefficient ($\epsilon$) = 0.2 \\
            & Entropy coefficient ($c_{\text{entropy}}$) = 0.01 \\
            & Value function coefficient ($c_{\text{vf}}$) = 0.5 \\
            & Maximum gradient norm = 0.5 \\
            & Learning rate = 0.00025 \\
        \midrule

        \textbf{Device parameters:}  & Acceleration constant $(\gamma)=1$\\
        & Inverse coherence time $(\alpha) = 0.02$ \\
        & Atom loss parameter $(\beta) = 0.02$ \\
        & Inter-zone transfer time $(T_G) = 10$\\
        & Lattice spacing of the grid $= 1$ \\
        \midrule
        \textbf{Other parameters} 
            & Window size ($W$) = 2 \\
            & Horizon length ($K$) = 5 \\
        \bottomrule
    \end{tabular}
    \label{tab:model-params}
\end{table*}

\newpage
\subsection{Size of the grids}

\begin{table}[ht]
\centering
\caption{The list of the size of the grids in each benchmark experiment. The benchmark 'transfer' represents the experiment in Section~\ref{section:transferability}.}
\begin{tabular}{ccc}
\toprule
Benchmark & Qubits  & Grid size (row $\times$ col) \\
\midrule
qaoa & 14 & $4 \times 10$ \\
groundstate & 14 & $4 \times 10$ \\
random & 30 & $7 \times 10$ \\
qnn & 50 & $5 \times 20$ \\
qft & 100 & $10 \times 20$ \\
hubbard & 100 & $10 \times 20$ \\
transfer-learning & 40-50 & $5 \times 20$ \\
transfer-learning & 80-100 & $10 \times 20$ \\
\bottomrule
\end{tabular}
\end{table}
\clearpage
\end{widetext}

\end{document}